\documentclass[preprint,authoryear, 3p]{elsarticle}

\usepackage{amssymb}
\usepackage{graphicx,subfigure}
\usepackage{algorithm}
\usepackage{amsmath}
\usepackage{amsthm}
\usepackage{enumerate}
\usepackage{tabto}
\usepackage{algpseudocode}

\begin{document}

\begin{frontmatter}



\title{A new structural stochastic volatility model of asset pricing\\and its stylized facts}


\author{Radu T. Pruna, Maria Polukarov, Nicholas R. Jennings}
\address{School of Electronics \& Computer Science, University of Southampton, UK}

\begin{abstract}
Building on a prominent agent-based model, we present a new structural stochastic volatility asset pricing model of fundamentalists vs. chartists where the prices are determined based on excess demand. Specifically, this allows for modelling stochastic interactions between agents, based on a herding process corrected by a price misalignment, and incorporating strong noise components in the agents' demand. The model's parameters are estimated using the method of simulated moments, where the moments reflect the basic properties of the daily returns of a stock market index. In addition, for the first time we apply a (parametric) bootstrap method in a setting where the switching between strategies is modelled using a discrete choice approach. As we demonstrate, the resulting dynamics replicate a rich set of the stylized facts of the daily financial data including: heavy tails, volatility clustering, long memory in absolute returns, as well as the absence of autocorrelation in raw returns, volatility-volume correlations, aggregate Gaussianity, concave price impact and extreme price events. 

\end{abstract}


\end{frontmatter}


\section{Introduction}\label{Section: Introduction}
\noindent A financial market is a broad term describing a marketplace where buyers and sellers participate in the trade of various financial assets at prices that reflect supply and demand. One of the most influential ideas in modern financial markets is the Efficient Market Hypothesis (EMH) proposed by \cite{fama1970efficient}. It postulates that prices fully reflect all available information, meaning that prices should adjust instantly and correctly to reflect new information. Consequently, all currently available relevant information regarding a particular asset is already incorporated in the market price. Only new information can change the price and it would be an immediate reaction of the market triggered by its arrival.

As a consequence of the random information arrival process, a market with such information efficiency leads to random sequences of price changes that cannot be predicted. Mathematically, it is equivalent to saying that prices follow martingales, where knowledge of past events never helps predict future values. It means that no profits can be generated from information-based trading since such earnings must have already been captured. According to \cite{lucas1978asset}, prices fully reflect all available information and follow martingales when all investors have rational expectations. 

Nevertheless, many theoretical and empirical implications of the EMH have been tested over the years. A range of studies raised several questions regarding the basic doctrines of the efficient markets model, especially on its view for asset price dynamics. The attacks on both theoretical and empirical sides of efficient market ideas developed and continued during the 1980s and 1990s with the work of \cite{grossman1980impossibility}, \cite{kindleberger1978manias} \cite{brock1992simple}, \cite{mehra1988equity} and \cite{hansen1983stochastic}, among others.

Most of the findings and debates regarding market efficiency and rationality are still unresolved, suggesting that markets may have non-trivial internal dynamics \citep{hommes2006heterogeneous}. Full rationality requires all agents to have knowledge about the beliefs of all the other participants in the market, thus making it impossible in a heterogeneous framework. On the other hand, heterogeneity obviously complicates the market models, making their analytical tractability almost impossible. A computational approach is therefore better suited for investigating the heterogeneous world, offering a strong motivation for developing agent-based models of financial markets \citep{chen2012agent}. 

Generally speaking, agent-based models are systems in which a number of heterogeneous agents interact with each other and their environment. In particular, agent-based simulations dealing with financial markets are known as Agent-Based Computational Finance Models \citep{tesfatsion2002agent}. Specifically, the financial markets can be viewed as examples of evolutionary systems populated by agents with different trading strategies. Recently, we have witnessed a transition in economics and finance from the rational agent approach to agent-based markets populated by heterogeneous agents with bounded rationality \citep{hommes2006heterogeneous, lux2008stochastic, chen2012agent}. This comes as an alternative to the rational expectations framework where all agents have fully, unbounded rationality.

In this context, the term bounded rationality, first introduced by \cite{simon1955behavioral}, suggests that individuals are not capable of the optimization levels required by full rationality. Instead, because of their limitations, market participants make choices that are satisfactory, but not necessary optimal. This means that agents usually follow strategies that have performed well in the recent past, according to different measures of accumulated wealth or profits. Interestingly, even the simplest trading strategies may survive the competition because of the co-evolution of prices and beliefs in the heterogeneous world.

One of the rapidly growing research directions in the field is to combine the agent-based modelling with econometrics and numerical simulations \citep{chen2012agent}. Nowadays, researchers do not only try to illustrate the basic mechanisms, but also quantitatively recreate the statistical properties of financial markets. Specifically, heterogeneous Agent Based Models (ABM) that use simple trading strategies have successfully generated a rich set of key properties that match the dynamics of asset prices in real financial markets. The ABM can provide useful insight on the behaviour of individual agents and also on the effects that emerge from their interaction, making the financial markets a very appealing application for agent-based modelling. This has led to the development of various models that aim to understand the links between empirical regularities of the markets and the complexities of the entire economic system (for recent surveys of this research area see e.g. \cite{chen2012agent, chiarella2008heterogeneity, hommes2006heterogeneous, lebaron2006agent, westerhoff2010simple}). However, while some models have been able to provide possible explanations for various properties of financial markets, no single one has produced or explained all the important empirical features of trading data, including volume, duration, price and especially asset returns \citep{lux2008stochastic, tesfatsion2002agent}.

To this end, independent studies have revealed a set of statistical properties common across different financial instruments, markets and time periods. Due to their robustness across various financial markets, they are called stylized facts in the econometrics literature \citep{cont2001empirical}. Specifically, these stylized facts target a wide range of aspects in the market, from price series and returns, to trading volume and volatility. The heterogeneous agent models attempt to explain these statistical properties of financial time series endogenously, by considering the interaction of market participants \citep{lux1995herd, lux1998socio, brock1998heterogeneous, lebaron1999time}. The most important stylized facts refer to the heavy tails in the distribution of asset returns, the persistence and clustering in return volatility, the characteristics of trading volume and the cross correlations between returns, volumes and volatility \citep{cont2001empirical}. Models which are able to produce outcomes close to the empirical stylized facts are very useful for studying the effects of regulations, trading protocols or clearing mechanisms. For this reason, the financial markets have become one of the most active research areas in agent-based modelling. 

Following the classification of \cite{chen2012agent}, there are, by and large, two main design paradigms of financial agents. The first one is referred to as the N-type design, and considers a relatively small number of agents with simple interactions, where all the agents' types and rules are given at the beginning of the design. On the other hand, in the autonomous-agent design complex interactions may lead to a large number of agents. Unlike the N-type design, in the autonomous-agent the agents are no longer divided into a fixed number of different types, their behaviour being customised via artificial intelligence algorithms. However, we are ultimately interested in constructing a model that mimics the most important properties of financial data and which is analytically tractable at the same time. Due to the high complexity of real markets it is hard to find the origins of critical events. Since we encounter a wide variety of trading strategies, motives and irrationality in the real world, a simple N-type design model is preferred. Focusing on how agents behave, we can observe the minimum conditions required to replicate the stylized facts in terms of both heterogeneity and complexity of the rules. Specifically, in this work we will focus on a two-type design, a particular case of the N-type design. Highly motivated and backed by empirical work, we will show how this type of financial model successfully meets our goals. In the same time, we will keep a high degree of transparency and clarity at the system level, making it easier for analysing the results.

In more detail, the main objective of Agent-Based Computational Finance Modelling is to propose an alternative to the apparent randomness of financial markets, trying to explain the most important properties of financial data. In particular, we are interested in simple structures that can reproduce the empirical findings to a high degree and which are quantitatively close to the real ones. Models that rely on simple trading strategies have proven themselves very efficient in generating important dynamics of real financial markets. Specifically, the fundamentalist vs. chartist models (to be defined) that incorporate all the important mechanisms of real financial markets are capable of recreating their key properties and are simple enough for analysis and computation. This strongly motivates our choice of focusing on fundamentalists vs. chartists model and we will show how effective it is in matching a rich set of stylized facts. 

In these models, the first type of agents, \textit{fundamentalists}, act as a stabilizing force in the market. Fundamentalists base their trading strategies and expectations of future prices on economic and market fundamentals such as earnings, growth and dividends. They believe in the existence of a fundamental price and invest relatively to its value. Therefore, a fundamentalist will most likely buy an asset that is undervalued; that is, one whose price is below the believed fundamental value, and sell an asset that is overvalued, that is, one whose price is above the fundamental value. In their opinion, the price series will eventually move or converge towards a long term equilibrium or fundamental value. 

In contrast, the other type of agents, \textit{chartists} or technical analysts, forecast the future prices entirely by modelling historical data. Surveys that brought to the attention of academics the use of technical analysis were conducted by \cite{frankel1987short, frankel1990exchange, frankel1990chartists}, \cite{allen1990charts, taylor1992use}, and more recently by \cite{gehrig2004use} and \cite{menkhoff2007obstinate}. Technical analysts do not take into consideration the market fundamentals and base their decisions entirely on observed historical patterns in past prices. They clearly believe that price developments display recurring patterns. Therefore, while the fundamentalists take into consideration only the fundamental price, the chartists base their decisions and trading strategies on historical prices.

Following some of the well-known approaches in the literature, we consider the Structural Stochastic Volatility model (FW) introduced by \cite{franke2012structural}. It is one of the most successful models in capturing the empirically observed traders' behaviour \citep{barde2015direct}. However, the price series generated by the FW model violates one of the core properties of financial time series \textendash \ its non-stationarity. We overcome this problem by extending the original model and changing the motion of the fundamental value over time. As a result, we drastically reduce the non-stationarity of the price series generated. Furthermore, a (parametric) bootstrap method is for the first time used to estimate the model's parameters in this setting, thus overcoming the joint-point problem associated with other methods. Finally, we evaluate the model and show it is able to match a rich set of stylized facts of real financial markets. Specifically, we make the following contributions:

\begin{itemize}
	\item Checking for stationarity in the price series generated by the FW model, we observe that the unit root test is rejected in more than one in every four simulations (see Section \ref{Fundamental Value} for more details). This is equivalent to having a price distribution with a mean and variance that do not change over time. In such a setting, we usually observe a mean-reverting behaviour with the price fluctuating around the fundamental value and which can easily be exploited. This strongly contradicts the well-known fact that in financial markets the price series are non-stationary. A non-stationary process has mean and variance that change over time and, as a rule, is unpredictable and cannot be forecasted.  
	
	The failure of simulations to produce the non-stationary prices observed in real financial markets is due to the unrealistic assumption of a constant fundamental value. Therefore, we build on the original model by making a novel change in the motion of the fundamental value over time. Accordingly, we assume the fundamental value obeys a random walk. Specifically, we set the fundamental price to follow a Geometric Brownian Motion (GBM), which is the most widely used model of describing stock price behaviour \citep{hull2006options}. As a consequence, the price series generated by the modified model (FW+) rejects the unit root test for stationarity in less than 5\% of the simulations.
	
	\item In the FW+ model, the agents interact stochastically, their switching between strategies being set by a discrete choice approach, influenced by a series of factors such as herding and price misalignment. In this setting, we estimate the model's parameters numerical values using a powerful method of simulated moments introduced by \cite{franke2011simple}. The method, based on a (parametric) bootstrap procedure used to evaluate the model's goodness-to-fit, is different from the other bootstraps or Monte Carlo experiments discussed in the literature and overcomes the joint-point problem (see Section \ref{Model Estimation} for more details). This is the first time it is applied to a model where the interaction between agents is based on a discrete choice approach. 
	
	\item One of the main objectives of agent-based financial modelling is to recreate the most important properties of financial data. An in depth analysis of the time series generated by our FW+ model is undertaken, providing a wide range of tests and arguments that reinforce the presence of a rich set of stylized facts including: lack of predictability and non-stationarity, absence of autocorrelation in raw returns, fat tails, volatility clustering and long memory in absolute returns, volume-volatility relations, aggregate gaussianity, price impact and extreme price events. To date, this is the only model that is reported to match such a rich set of the stylized facts of real financial markets. Therefore, we illustrate one of the essential aims of the agent-based financial modelling. Namely, that many of the stylized facts arise and can be explained just from the interaction of market participants. 
\end{itemize}

The remainder of the paper is organised as follows. In Section \ref{Model Definition} we formally define the fundamentalist vs. chartist asset pricing model and its key mechanisms, discussing its dynamic properties. In Section \ref{Model Estimation} we estimate and validate the model's parameters. Next, in Section \ref{Results}, we present our results. We give a thorough analysis of the price series generated by the agent-based model, integrating a wide range of tests and evidence that demonstrate the presence of a rich set of stylized facts. Section \ref{Conclusion} concludes. 

\section{Model formulation}
\label{Model Definition}

\noindent In this section, we extend the fundamentalists vs. chartists asset pricing model by \cite{franke2012structural} (FW) to capture new properties usually observed in real financial markets. In Sections \ref{Price adjustments} and \ref{Evolution of the market shares} we formally define the asset pricing model, with its pricing dynamics and agents' interactions. Next, in Section \ref{Fundamental Value}, we describe our contribution regarding the movement of the fundamental value, finishing with a discussion on the dynamics of the model in Section \ref{Model dynamics}.

\subsection{Price adjustments}
\label{Price adjustments}

\noindent The asset price changes are determined by excess demand, as in the original FW model, following some of the most prominent examples in the literature \citep{beja1980dynamic, farmer2002price}. Here, by excess demand we mean the precise positive or negative orders per trading period. The specific demand of each trader type is kept as simple as possible, in the form of the demand per average trader. For fundamentalists, the demand is inversely related to the difference between the current price and the fundamental value. That is, at time $t$ their core demand $D_t^f$ is proportional to the gap $(p_t^f-p_t)$, where $p_t$ is the log price of the asset at time $t$, while the $p_t^f$ is the fundamental log value at time $t$. Similarly, the core demand of the chartists' group, $D_t^c$, is proportional to the price changes they have just observed, $(p_t-p_{t-1})$, where $p_t$ and $p_{t-1}$ are the log prices at time $t$ and $t-1$, respectively. 

The wide variety of within-group specifications are captured by noise terms added to each of the core demands. These terms encapsulate the within-group heterogeneity and scale with the current size of the group. Specifically, the noise is represented by two normally distributed random variables $\epsilon_t^f$ and $\epsilon_t^c$ for fundamentalists and chartists, respectively. In other words, one can think of the noise variables as a convenient way of capturing the heterogeneity of markets populated by hundreds or thousands of different agents. Each of the two noise terms are sampled at every iteration and added to the deterministic part of the demand, leading to the total demand per agent within the corresponding group. 

Thus, combining the deterministic and stochastic elements, we get the net demand of each group for the asset in period $t$ as follows:
\begin{equation}\label{eq: 1}
D_t^f=\phi(p_t^f-p_t)+\epsilon_t^f \quad \quad \epsilon_t^f \sim  \mathcal{N}(0,\sigma_f^2) \quad \quad \phi > 0, \\\ \epsilon_f>0,
\end{equation}
\begin{equation}\label{eq: 2}
D_t^c=\chi(p_t-p_{t-1})+\epsilon_t^c \quad  \quad \epsilon_t^c \sim  \mathcal{N}(0,\sigma_c^2) \quad  \quad \chi \ge 0, \\\ \epsilon_c\ge 0,
\end{equation}
where $\phi$ and $\chi$ are constants denoting the aggressiveness of traders' demand and $\sigma_t^f$ and $\sigma_t^c$ are noise variances.

The agents are allowed to switch strategies at each iteration, so their market fractions fluctuate over time. For simplicity, we fix the agents' population size at $2N$. Let $n_t^f$ and $n_t^c$ be the number of fundamentalists and chartists in the market at time $t$, respectively. We define the \textit{majority index} of the fundamentalists as: 
\begin{equation} \label{eq:3.3}
x_t = (n_t^f-n_t^c)/2N
\end{equation}

Therefore, $x_t \in [-1,1]$, a value of $x_t=-1$  $(x=+1)$ corresponds to a market where all traders are chartists (fundamentalists). Moreover, the market fractions can be expressed in terms of the majority index as:
\begin{equation}\label{eq: 4}
n_t^f/2N=(1+x_t)/2, \quad \quad n_t^c/2N=(1-x_t)/2.
\end{equation}

The (scaled) total demand $D_t$, given by the equation: 
\begin{equation}\label{eq: 5}
D_t = n_t^fD_t^f + n_t^cD_t^c = (1+x_t)D_t^f/2 + (1-x_t)D_t^c/2
\end{equation}  
will generally be in disequilibrium. That is, the total demand of the agents will not add up to zero and we will have an excess of either supply or demand. Following the early examples in the literature \citep{farmer2002price}, a market maker is assumed to absorb the excess supply and provide any excess demand. The market maker mediates the transactions between investors and provides liquidity. He sets the price by supplying stock out of its inventory and raising the price if there is excess demand, while accumulating stock and lowering the price when there is excess supply. Specifically, the market maker reacts to the imbalance between demand and supply by proportionally adjusting the price with a constant factor $\mu >0$. 

Accordingly, the equation determining the price for the next period $t+1$ results from equations \ref{eq: 1}-\ref{eq: 5} as:
\begin{equation}\label{eq: 6}
p_{t+1} = p_t + \frac{\mu}{2}[(1+x_t)\phi(p_t^f-p_t) + (1-x_t)\chi(p_t-p_{t-1})+\epsilon_t],
\end{equation}  
\begin{equation}\label{eq: 7}
\epsilon_t\sim\mathcal{N}(0,\sigma_t^2), \quad \quad \sigma_t^2=[(1+x_t)^2\sigma_f^2+(1-x_t)\sigma_c^2]/2.
\end{equation}  

Equation \ref{eq: 7} is derived as the sum of the two normal distributions from Equations \ref{eq: 1} and \ref{eq: 2}, multiplied by the market fractions $(1\pm x_t)/2$. The result is a new normal distribution with mean zero and variance equal to the sum of the two single variances. The combined variance $\sigma_t$ depends on the variations of the market fractions of the fundamentalists and chartists. This random time-varying variance is a key feature of the model, being termed the \textit{structural stochastic volatility} (SSV) of returns (defined as the log differences in prices) in \cite{franke2009validation}. 

\subsection{Evolution of the market shares}
\label{Evolution of the market shares}

\noindent To complete the model, it remains to set up the motions of the market fractions $n_t^f$ and $n_t^c$. They are predetermined in each period and change only from one period to the next one. Following the discrete choice approach (DCA) introduced by \cite{brock1997rational}, the two market shares $n_{t+1}^s$ $(s=f,c)$ can be determined using the multinomial logit model. In the basic setting, some payoff indices $u_t^c$ and $u_t^f$ are considered, usually derived from past gains of the two groups. The market fractions can be expressed as $n_{t+1}^s=exp(\beta u_{t}^s)/exp(\beta u_{t}^f)+exp(\beta u_{t}^c)$, where $\beta$ is the intensity of choice. Dividing by $exp(\beta u_t^f)$, the market fraction of fundamentalists is given by $n_{t+1}^f=1/\{1+exp[-\beta (u_{t}^f-u_{t}^c)]\}$.

Note that the difference in any utility variables, $(u_t^f-u_t^c)$, can be viewed as a measure of relative attractiveness of fundamentalist trading. Hence, we may change the notation, making use of the\textit{ attractiveness level}, $a_t$, defined as the difference $(u_t^f-u_t^c)$. The discrete choice approach is then given by:
\begin{equation}\label{eq: 8}
n_{t+1}^f=\frac{1}{1+exp(-\beta a_t)}, \quad n_{t+1}^c=1-n_{t+1}^f.
\end{equation}

We normalize all the demand terms in the price rule \ref{eq: 6} by using a market impact $\mu=0.01$. Furthermore, we fix the intensity of choice $\beta=1$. Of course, setting these values is just a matter of scaling the market impact on prices and the relative attractiveness level $a_t$ of fundamentalism, respectively. 

Note that the market fractions are directly influenced by the attractiveness level. An increase in the index $a_t$ leads to an increase in the market share of the fundamentalists. For this reason, it is extremely important to define the exact mechanism of the attractiveness level and all of its components. 

Following the FW model setting, we consider three principles that may influence the way in which the agents choose one of the two strategies. The first principle is a herding mechanism, meaning that the more agents are in a group the more attractive that group becomes. This idea has been long used in the literature (see, for example, \cite{kirman1993ants} and \cite{lux1995herd}) and can be easily represented by a term proportional to the most recent difference in the market fractions $(n_t^f-n_t^c)$.

The second principle is based on a certain predisposition towards one of the two trading strategies. This can be directly captured by a constant $\alpha_0$, which is positive (negative) if the agents have a priori preference for fundamentalism (chartism). Finally, the third component encapsulates the idea of price misalignment. This is empirically backed by \cite{menkhoff2009heterogeneity}, who observes that when the price is further away from its fundamental value professionals tend more and more to anticipate its mean reversion toward equilibrium. In a setting where the current price tends to move far away from the fundamental value, chartism becomes riskier and the fundamentalism is more attractive. Hence, it is convenient to make $a_t$ rise in proportion to the squared deviations of the price $p_t$ from the fundamental value $p_t^f$. 

Combining the three components of the attractiveness level, we have:
\begin{equation}\label{eq: 9}
a_t=\alpha_0 +\alpha_n(n_t^f-n_t^c)+\alpha_p(p_t-p_t^f)^2,
\end{equation}   
where $\alpha_0$ is the predisposition parameter, $\alpha_n>0$ captures the herding parameter and $\alpha_p>0$ measures the influence of price misalignment. 

In summary, the model is governed by two central dynamic equations. Firstly, the price adjustments are made according to Equations \ref{eq: 6} and \ref{eq: 7}, with the structural stochastic component $\sigma_t$. Secondly, the changes of the market fractions $n_t^f$ and $n_t^c$ are described by Equations \ref{eq: 8} and \ref{eq: 9}, encapsulating a herding mechanism corrected by a strong price misalignment. The whole system has a recursive structure that is easily forward iterated. 

\subsection{Motion of the fundamental value}
\label{Fundamental Value}
\begin{figure}
	\centering
	\includegraphics[width=0.63\textwidth]{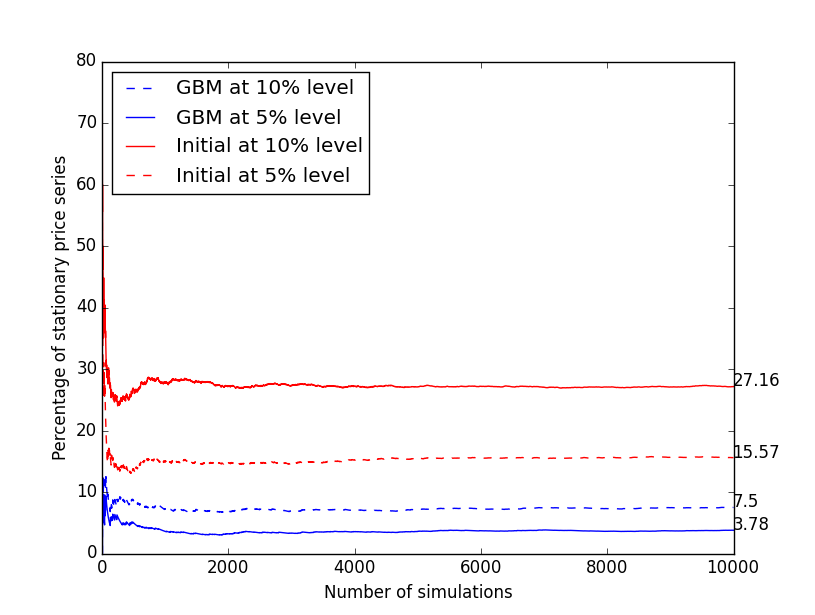}
	\caption{Augmented Dickey-Fuller stationarity test results.}
	\label{fig:Stationarity}
\end{figure}

\noindent Financial time series are characterised by the lack of predictability, mathematically known as the martingale property. It is a fundamental ingredient of theoretical models stating that knowledge of past events never helps predicting the future movements. The lack of predictability of price series is well known and documented, with a well-established body of literature offering plausible generic explanations of this stylized fact \citep{lux2008stochastic}. It is explained by traditional finance as a consequence of the informational efficiency, that is, all currently available relevant information is already embedded in the market price. Hence, the price can be changed only by the arrival of new information.  

A similar, very common property of financial time series data is its non-stationarity. In particular, a non-stationary process is a stochastic process whose joint probability distribution changes when shifted in time. Consequently, its mean and variance change over time. Usually, stock prices are defined as examples of random walks, a non-stationary process. It is commonly assumed that non-stationary data is unpredictable and cannot be forecasted.

In more detail, the non-stationarity of financial data has been discussed and studied for a long time and, similarly to the martingale property, arises from the theory of efficient markets \citep{pagan1996econometrics}. A property associated with non-stationarity is the unit root, which states that one is not able to reject the hypothesis that the time series follows a random walk or a martingale process. Alternatively, the series is said to be integrated of order one. A huge body of literature has been spawned on the question of unit roots, resulting in a wide range of statistical tests that can be applied to financial data. 

One of the widely used tests of non-stationarity is the Augmented Dickey Fuller (ADF) unit root test. Following this approach, we apply the classical ADF test (allowing for a constant and trend order) to the FW model, where the fundamental value is kept constant, $p_t^f=0$. In 10000 simulations, the unit root test was rejected 2716 times with the p-value being less than the critical value at 10\%. Moreover the test was rejected 1557 times with the p-value less than the critical value at 5\%. The p-values are obtained using the updated tables from \cite{mackinnon2010critical}. Therefore, we can say that the price series generated by the model using a constant fundamental value are stationary approximately 27\% of the time, which strongly contradicts the behaviour of real financial time series. 
\begin{table}
	\centering
	\caption{Model parameters}
	\label{table: 1}
	\begin{tabular}{lllllllll}
		\hline
		$\phi$ & $\chi$ & $\sigma_f$ & $\sigma_c$ & $\alpha_0$ & $\alpha_n$ & $\alpha_p$ & $\mu_p$ & $\sigma_p$ \\ 
		\hline
		0.121 & 1.555 & 0.592 & 1.917 & -0.301 & 1.990 & 22.741 & 0.01 & 0.157\\
		\hline
	\end{tabular}
\end{table}
The failure to pass the unit root test is mainly due to the unrealistic assumption of a constant fundamental value. As a solution, we extended the setting by allowing the fundamental value $p_t^f$ to change over time. Specifically, we set $p_t^f$ to follow a geometric Brownian motion. This is the most widely used model of describing stock price behaviour \citep{hull2006options} and is usually applied in quantitative finance.  Mathematically, the fundamental price is given by 
\begin{equation}
dp_t^f=\mu_pp_t^fdt+\sigma_pp_t^fdW_t,
\end{equation}
where $W_t$ is a Weirner process, $\mu_p$ is the percentage drift and $\sigma_p$ is the percentage volatility. The exact values of the drift and volatility will be estimated together with all the other model's parameters in Section \ref{Model Estimation}. 
\begin{figure}
	\centering     	
	\subfigure[Simulated log prices]{\includegraphics[width=\textwidth]{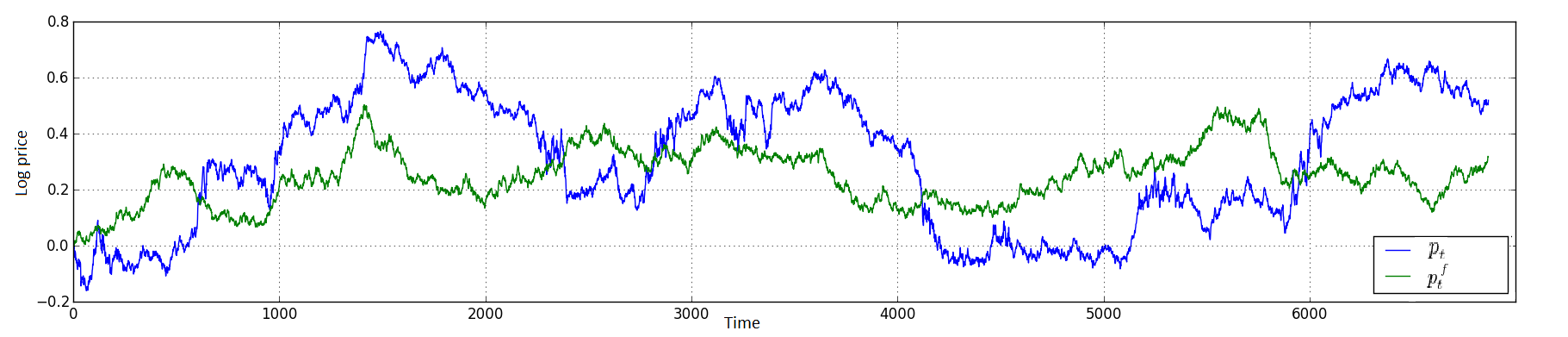}}
	\subfigure[Simulated majority index]{\includegraphics[width=\textwidth]{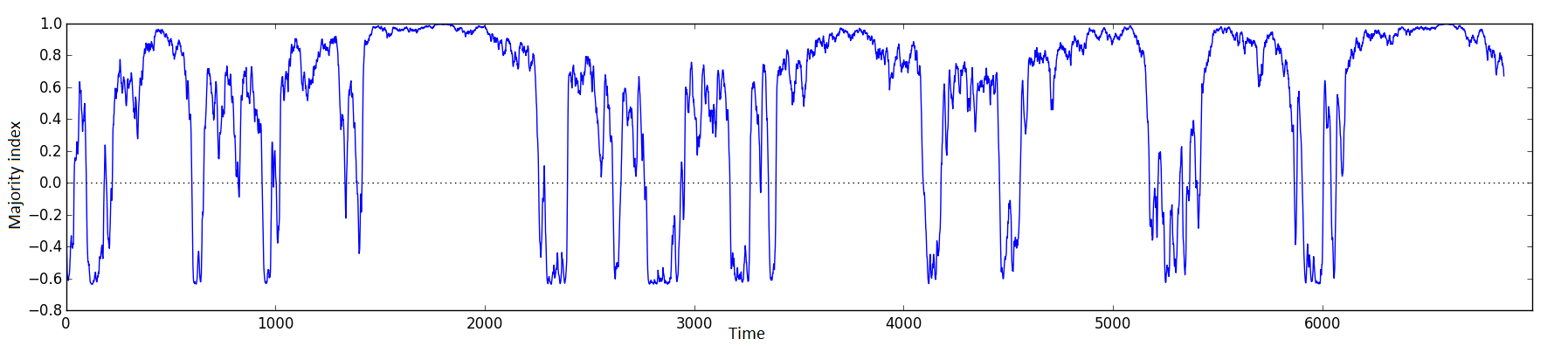}}
	\subfigure[Simulated returns]{\includegraphics[width=\textwidth]{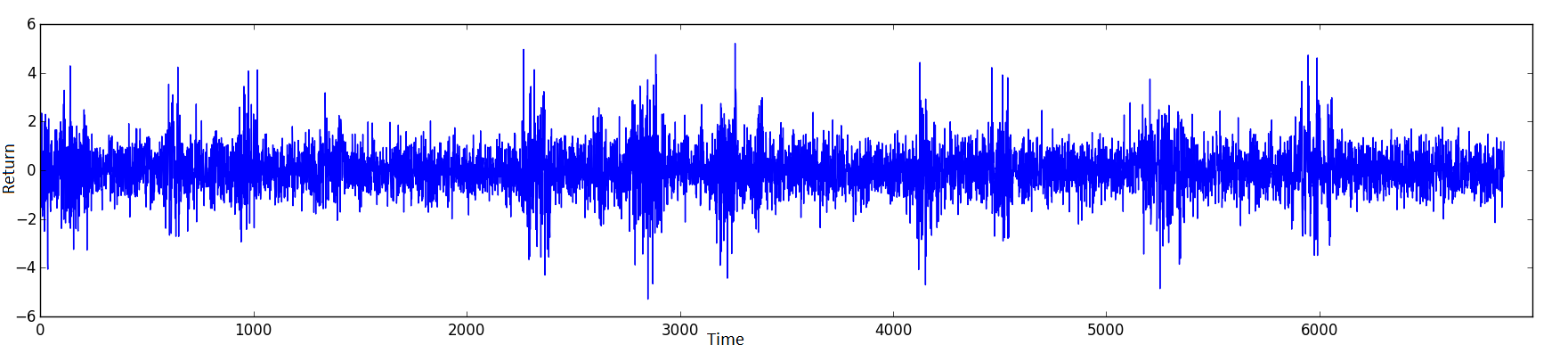}}
	\subfigure[Empirical returns]{\includegraphics[width=\textwidth]{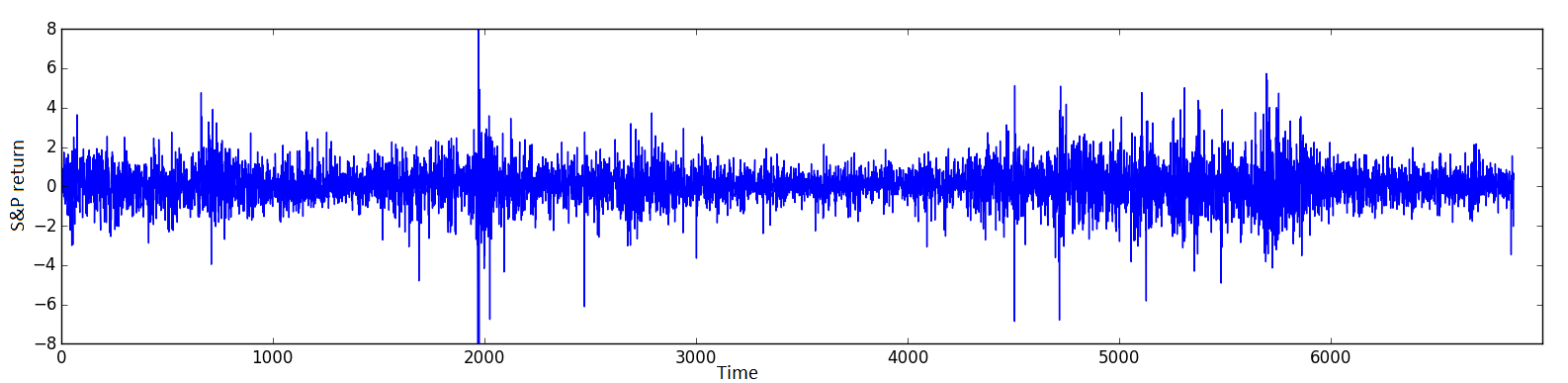}}
	\caption{Simulated price series ($p_t$), fundamental price series ($p_t^f$), majority index ($x_t$) and returns ($r_t$) over 6867 time periods, together with empirical S\&P 500 returns.}	
	\label{fig:simulation}
\end{figure} 
With this change in the model, we run the Augmented Dickey Fuller test once again over 10000 different simulations. Now, the test is rejected only 750 times with the p-value less than the critical value at 10\% and 378 times with the p-value less than the critical value at 5\%. Thus, we can say that the price series generated by the new model (FW+) are non-stationary in more than 93\% of the simulations (with 95\% confidence interval). A visual representation of the number of non-stationary price series generated in both models can be seen in Figure 1. We plot the percentage of the non-stationary price series generated as we increase the number of simulations. The red and blue lines represent the initial and our modified model, respectively. The hard (dotted) lines are the computed ADF results at 10\% (5\%) critical values. We can observe how the number of non-stationary price series is drastically reduced, from 27\% with the initial model (at 10\% critical size) to 4\% in our modified model (at 5\% critical size).

Therefore, we can say that the new model produces more realistic price series, since in real financial markets the prices follow martingales where knowledge of past events does not help predict the mean of the future changes.  Unlike the original (FW) model, our improved (FW+) setting better reflects some of the fundamental properties of real financial time series \textendash \ their non-stationarity and unpredictability.

\subsection{How the model functions}
\label{Model dynamics}

\noindent We now turn to explore the qualitative measures that demonstrate the existence of the most important stylized facts of financial markets. An in-depth analysis of the price, returns, volatility and volume series generated by the FW+ model will be performed in Section \ref{Results}. The numerical parameters used for all the simulations are given in Table \ref{table: 1}.  

In Figure \ref{fig:simulation} we present a simple run of the model over 6867 days, covering the same time span as the empirical data represented by Standard and Poor (S\&P) stock market index from January 1980 to mid-March 2007. The top panel illustrates the (log) price series $p_t$ generated by the model and the fundamental values $p_t^f$. We can observe the irregular swings in the prices, with a considerable amplitude, similar to the behaviour of real financial series, reflected by the empirical returns in Figure \ref{fig:simulation}(c). The second panel shows the corresponding composition of the traders, or the fluctuations in the market fraction. The majority index $x_t$ is moving continuously, changing from periods of fundamentalists domination (positive values) to periods of chartists domination (negative values). However, it shows that most of the time the market is dominated by fundamentalists ($x_t>0.6$), with sudden swings to chartism ($x_t\approx-0.65$). 

A quick visual inspection of simulated price series and majority index plotted in Figures \ref{fig:simulation} (a) and (c), respectively, reveals a high market fraction of fundamentalists in the presence of strong mispricing. On the other hand, the chartists increase in number when the price gets closer to the fundamental value. This behaviour is obvious from the model definition. The mispricing element in the attractiveness level equation makes fundamentalism more appealing when the price drifts from the fundamental value. However, an increasing number of fundamentalists begin to dominate the market and their actions start the mean reverting process of the price towards the fundamental value. As the current price gets closer to the fundamental value, fundamentalism is no longer attractive or profitable, making the agents switch to chartism. The herding element takes over, leading to a majority fraction of chartism, whose actions will move the price away from the fundamental value. 

The third panel in Figure \ref{fig:simulation} demonstrates the implications of these irregular switches in the agents' strategies on returns. Since the chartists have a greater variability in demand, $\sigma_c^2>\sigma_f^2$, by comparing Figures \ref{fig:simulation} (b) and (c) we observe that the level of returns during a chartism domination exceeds its level in a fundamentalism regime. Therefore, it appears that normal sequences of returns are interrupted by outbursts of increased volatility, when the majority of agents are chartists.  

\section{Estimation of the model's parameters}
\label{Model Estimation}

\noindent In this section we discuss the formal numerical estimation of the model's parameters, using one of the most prominent methods from the literature, the Method of Simulated Moments (MSM). For an overview of the method of simulated moments and its applications to agent-based models see the work of \cite{winker2001indirect, gilli2003global, amilon2008estimation, franke2009applying}. This technique is based on an objective function that is optimised across the set of FW+ model's parameters ($\phi, \chi, \sigma_f, \sigma_c, \alpha_0, \alpha_n, \alpha_p, \mu_p, \sigma_p$). The method refers to a set of statistics, also known as moments, arising from the simulations. The basic idea is that these moments should be close to their empirical counterparts, with the distance between them being captured by the objective function. An agent-based model is not supposed to mimic the exact economy or financial markets, but rather it should explain some of the most important stylized facts of financial markets, making this estimation method suitable for our purposes. 
\begin{algorithm}
	\caption{Bootstrap estimation of the covariance matrix}
	\begin{algorithmic}	
		\State Set $I_0 = \{1,2,\dots,T\}$
		\For{$b = 1:B$}
		\State Draw T random numbers with replacement from $I_0$
		\State Construct bootstrap sample $I_b = \{t_1^b,t_2^b,\dots,t_T^b\}$
		\State Calculate vector of moments $m_b=(m_1^b,\dots,m_9^b)$ from $I_b$
		\EndFor
		\State $\bar{m} = \frac{1}{B} \sum_{b} m_b$
		\State $\hat{\Sigma} = \frac{1}{B} \sum_{b} (m_b-\bar{m})(m_b-\bar{m})'$
		\State $W=\hat{\Sigma}^{-1}$
	\end{algorithmic}
	\label{algo: estimation}
\end{algorithm}
In this paper, we calibrate the FW+ model using the same dataset of T=6866 daily observations of the Standard and Poor (S\&P 500) stock market index from January 1980 to mid-March 2007, as in the FW model. The estimation was conducted such that the four most discussed statistical properties of empirical financial data are matched. These include absence of autocorrelation in raw returns, heavy tails, volatility clustering and long memory. The returns, or (log) price changes, are expressed as percentage points, such as:
\begin{equation}\label{eq: 10}
	r_t=100 (p_t-p_{t-1}).
\end{equation}      
Given this, the volatility of returns is defined as their absolute value $v_t=|r_t|$.

In order to conduct the quantitative analysis, the four stylized facts were measured using a number of summary statistics, or moments. The first moment considered is the first-order autocorrelation coefficient of raw returns. It needs to be close to zero, so that it agrees with the empirical findings on this matter \citep{kendall1953analysis, fama1965behavior, bouchaud2003theory, chakraborti2011econophysics, cont2014price}. This will limit the chartists' price extrapolations of the most recent price changes. Furthermore, if the model generates coefficients that are insignificant, all the autocorrelations at longer time lags will vanish too. 

Furthermore, the remaining three moments are dealing with the volatility of returns. First of all, the model should suitably scale the overall volatility, thus limiting the general noise brought by the two variances $\sigma_f$ and $\sigma_c$. The mean value of the absolute returns is considered. Next, the heavy tail is measured by the Hill tail index of the absolute returns. The tail is specified as the upper 5\% in order to eliminate bias and consider a more accurate tail index.   

The long memory effects are captured by the autocorrelation function (ACF) of the absolute returns up to a lag of 100 days. In particular, the autocorrelation decays as we increase the lag, without becoming insignificant. The entire profile has to be matched and is sufficiently well represented by six different lag coefficients ($\tau=1, 5, 10, 25, 50, 100$). 
   
Thus, the model is evaluated on the basis of nine moments (the Hill estimator, volatility, first order autocorrelation of the raw returns and the autocorrelation at lags $\tau=1, 5, 10, 25, 50, 100$ for the absolute returns), summarized in a (column) vector $m=(m_1,...,m_9)'$ (the prime denotes transposition). Applying the method of simulated moments, they have to come as close as possible to their equivalent empirical moments, $m^{emp}$, calculated on the daily S\&P 500 stock market index. 

The distance between the two vectors $m$ and $m^{emp}$ is defined as a quadratic function with a suitable weighting matrix $W \in R^{9x9}$ (defined shortly). Hence, the distance is given by:
\begin{equation}\label{eq: 11}
	J=J(m,m^{emp};W)=(m-m^{emp})'W(m-m^{emp}).
\end{equation}

The weight matrix $W$ accounts for the moments' sampling variability, its determination being crucial for the model's parameters. The idea is that the higher the sampling variability of a given moment $i$, the larger the differences between $m_i$ and $m_i^{emp}$ that can still be considered insignificant. This behaviour can be achieved by correspondingly small diagonal elements $w_{ii}$. Moreover, the matrix $W$ should support possible correlations between single moments. One obvious choice for $W$ is the inverse of an estimated variance-covariance matrix $\hat{\Sigma}$ of the moments \citep{franke2009applying},
\begin{algorithm}
	\caption{Parameter estimation}
	\begin{algorithmic}
		\For{$a = 1:1000$}
		\State $\hat{\theta}^a =$ minimise(funJ, method=Nelder-Mead)	
		\EndFor
		\State $\hat{\theta}=\hat{\theta}^{\tilde{a}}$, where $\tilde{a}$ such that $\hat{J}^{\tilde{a}}$= median of $\{\hat{J}^a\}_{a=1}^{1000}$ and $\hat{J}^a=J[m^{{a}}(\hat{\theta};S), m_T^{emp}]$
		\State
		\State def funJ($\theta$):
		\State $\quad$		Simulate model using vector of parameter $\theta$
		\State $\quad$ 		Get simulated moments $m_{sim}=(m_1, \dots, m_9)$
		\State $\quad$ 		Get $J=J(m^emp,m_T^{emp};W)=(m^{sim}-m_T^{emp})'W(m^{sim}-m_T^{emp})$
		\State $\quad$ 		Return $J$
	\end{algorithmic}
	\label{algo: parameter}
\end{algorithm} 

\begin{equation}\label{eq: 12}
	W=\hat{\Sigma}^{-1}.
\end{equation}     

The covariances in $\hat{\Sigma}$ are estimated by a bootstrap procedure used to construct additional samples from the empirical observations. In the literature, this is often carried out by a block bootstrap \citep{winker2007objective, franke2011estimation, franke2012structural}. However, the original long-range dependence in the return series is interrupted every time two non-adjacent blocks are pasted together. Hence, the independence of randomly selected blocks cannot reproduce the dependence structure of the original sample. This is known as the joint-point problem \citep{andrews2004block}. Since our estimation is concerned with summary statistics, we can overcome the joint-point problem by avoiding the block bootstrap. 

Correspondingly, in order to estimate the variance-covariance matrix $\hat{\Sigma}$, we use a new bootstrap method first described by \cite{franke2011simple}. In their work, the authors apply this method to the FW model with the interaction between agents being modelled by the transition probability approach \citep{lux1995herd}. Therefore, we are for the first time applying this new bootstrap framework to a model where the agents interact stochastically according to the discrete choice approach \cite{brock1997rational}. Departing from the traditional block bootstrap, we sample single days and, associated with each of them, the history of the past few lags required to calculate the lagged autocorrelations. In more detail, the procedure is presented in Algorithm \ref{algo: estimation}. We construct the set of time indices, $I_0=\{ 1,2, \dots, T \}$, and for each bootstrap sample $b$, we can sample directly from it. Accordingly, a bootstrap sample is constructed by $T$ random draws with replacement from $I_0$. Repeating this $B$ times, we have $b=1,\dots,B$ index sets,
\begin{equation}\label{eq: 14}
I_b=\{ t_1^b, t_2^b, \dots, t_T^B \},
\end{equation}  
from witch the bootstrapped moments are obtained. 

For a good representation, the bootstrap method is repeated 5000 times, obtaining a distribution for each of the moments. Formally, let $m^b=(m_1^b,...,m_9^b)'$ be the resulting vector of moments, and $\bar{m}=(1/B)\sum_b m^b$ be their mean values. Then, estimate of the moment covariance matrix $\hat{\Sigma}$ becomes,
\begin{equation}\label{eq:3.14}
	\hat{\Sigma}=\frac{1}{B}\sum_{b=1}^B (m^b-\bar{m})(m^b-\bar{m})'.
\end{equation}   

Going back to the estimation problem, we are interested in the set of parameters that minimise the distance function $J$ from Eq. \ref{eq: 11}. In order to reduce the variability in the stochastic simulations, the time horizon is chosen to be longer than the empirical sample period $T$, commonly defined as $S=10\cdot T$. Repeated simulations over $S$ periods (or days) are carried out, in search of the set of parameters that minimise the associated loss. To this end, let $\theta$ be the vector of parameters and $m=m(\theta;S)$ denote the moments to which a vector $\theta$ gives rise. 

Furthermore, the comparability of different trials of $\theta$ is determined by a random seed $a=1,2,\dots$, let us say. Moreover, let $m^a(\theta,S)$ denote the moment vector obtained by simulating the model with a parameter vector $\theta$ over $S$ periods on the basis of a random seed $a$. The parameter estimates on a random seed $a$, denoted $\hat{\theta}^a$, are the solution of the following minimisation problem,
\begin{equation}\label{eq: 15}
	\hat{\theta}^a=\operatornamewithlimits{argmin}_{\theta}J[m^a(\theta;S),m_T^{emp}],
\end{equation}  
where $m_T^{emp}$ is the moment vector for the empirical S\&P 500 data\footnote{For the actual minimisation problem we use the Nelder-Mead simplex search algorithm \citep{nelder1965simplex}.}. 

Although one may think that a simulation over $S=68660$ days provides a large sample to base the moments on, the variability arising from such different samples is still considerable. Hence, it seems most appropriate to carry out a great number of estimations ($a=1000$) and choose the one with the median loss. The parameter set $\hat{\theta}$ giving this associated loss will be our representative estimation (see Algorithm \ref{algo: parameter}). Specifically, using Equation \ref{eq: 15} we have,
\begin{equation}\label{eq: 16}	
\begin{split}
	&\hat{\theta}=\hat{\theta}^{\tilde{a}}, where \: \tilde{a} \: is \: such \: that \: \hat{J}^{\tilde{a}} \: is \: the \: median \: of \: \{\hat{J^a}\}_{a=1}^{1000}, \: and\\ 
	&\hat{J^a} = J[m^a(\hat{\theta}^a;S), m_T^{emp}], \quad a=1,\dots,1000
\end{split}
\end{equation}    

\begin{figure}
	\centering  	
	\subfigure{\includegraphics[width=0.50\textwidth]{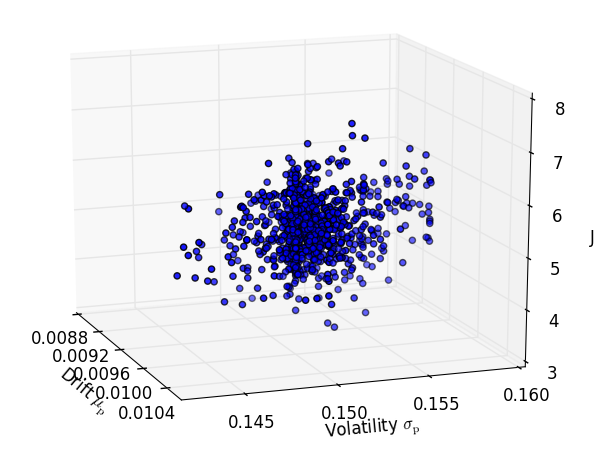}}
	\hfill
	\subfigure{\includegraphics[width=0.49\textwidth]{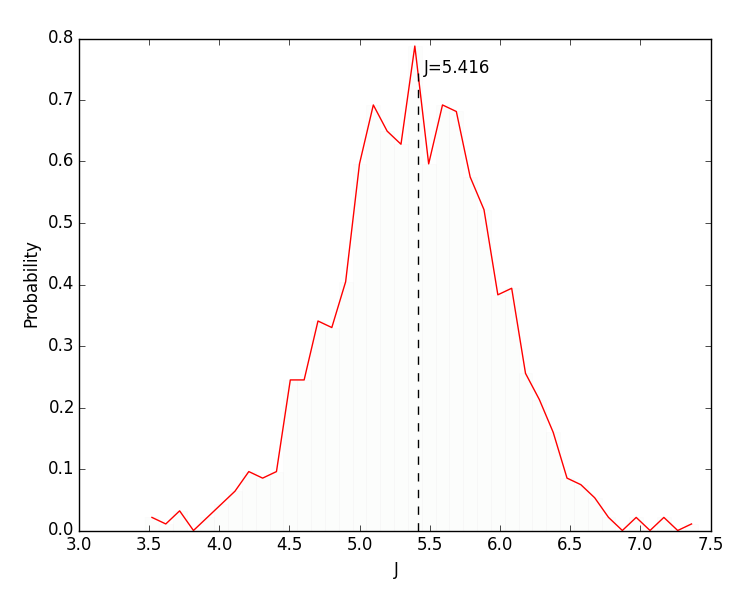}}	
	\caption{Distribution of objective function J.}	
	\label{fig:J_dist}
\end{figure}
The final parameter vector $\hat{\theta}$ resulting from the estimation has already been reported in Table \ref{table: 1}. The corresponding minimized loss is $5.416$,
\begin{equation}\label{eq: 17}	
\hat{J} = J[m^{\tilde{a}}(\hat{\theta};S), m_T^{emp}]=5.416
\end{equation}     

A visual representation of the distribution of the objective function $J$ can be observed in Figure \ref{fig:J_dist}. In the left panel, we plot the function as small changes of the last two parameters of the model, the drift and volatility of the fundamental price. We observe how it doesn't depart from small values, staying in the neighbourhood of the median we chose as optimal value. In the right panel, we plot the actual distribution of $J$.

To sum up, the estimation of the model's parameters is based on the minimisation of Equation \ref{eq: 15}, where the objective function $J$ is defined by Equations \ref{eq: 11}, \ref{eq: 12}, \ref{eq: 14} and the set of nine moments described earlier. The model was validated on the S\%P 500 data, leading to the set of parameters presented in Table \ref{table: 1}. These are the values of parameters we used to generate all the results discussed in the following section. 

\section{Results}
\label{Results}

\noindent In this section, we explore the statistical properties generated by our model. Specifically, we provide both quantitative and qualitative measures that demonstrate the existence of the most important stylized facts of financial markets. An in-depth analysis of the price, returns, volatility and volume series generated by the new model will be performed. We show that the model is able to match a rich set of properties including martingales, absence of autocorrelations in raw returns, heavy tails, volatility clustering and long memory in absolute returns, volume-volatility relations, aggregate Gaussianity, a concave price impact and extreme price events.   

\subsection{Absence of autocorrelations}
\label{autocorrelation}

\noindent The lack of predictability of price series is one of the most distinctive characteristics of financial time series. It has puzzled both researchers and investors over the years, making it one of the most discussed properties of financial data. A similar, very common property of financial data is its non-stationarity. Stock prices have probability distributions whose mean and variance change over time. Accordingly, they are defined as examples of random walks, a non-stationary process. This has motivated us to change the behaviour of the fundamental price to a geometric Brownian motion, one of our main contributions, such that the price series generated by the model are non-stationary (see Sec. \ref{Fundamental Value}). 

A further, well-known property of financial data states that price movements do not exhibit any significant autocorrelation \citep{cont2001empirical}. The autocorrelation function defined as:
\begin{equation}
C(\tau)=corr(r(t,\Delta t),r(t+\tau , \Delta t)),
\end{equation}
where $corr$ denotes the sample correlation, rapidly decays and even for small interval time  is close to zero. 

Many studies have found a rapid decline of ACF after the first lag, therefore confirming the absence of (linear) autocorrelation in returns at all horizons and making it a well accepted stylized fact \citep{working1934random, kendall1953analysis, fama1965behavior, bouchaud2003theory, chakraborti2011econophysics}.

\begin{figure}
	\centering
	\includegraphics[width=0.6\textwidth]{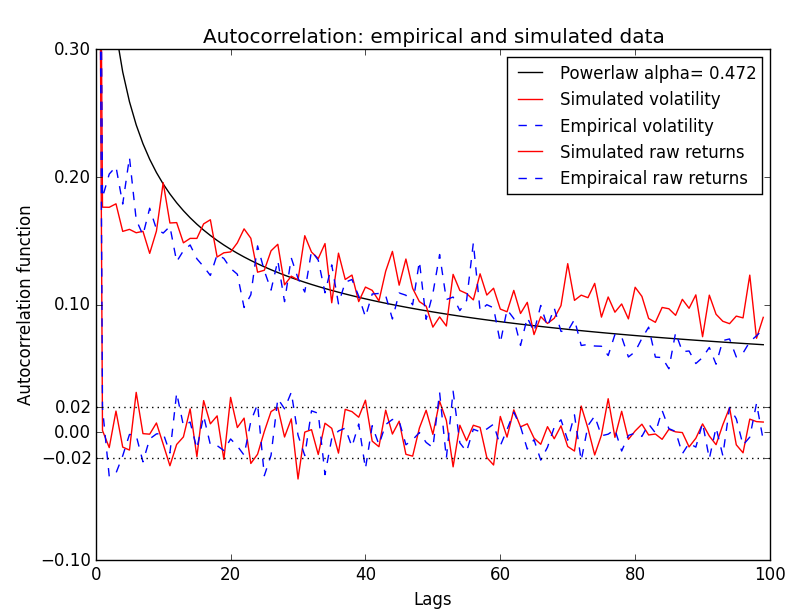}
	\caption{Autocorrelation function of returns. The red lines (blue dots) indicate empirical (simulated) returns. The two upper (lower) lines represent the autocorrelation of volatility (raw returns) at lags $\tau=1,...,100$.}
	\label{fig:acf}
\end{figure}

The absence of autocorrelation in returns can easily be demonstrated using the autocorrelation function. This basic stylized fact is demonstrated in Figure \ref{fig:acf}, where the ACF of raw returns is computed at lags from 1 to 100. We compare the ACF of both simulated and empirical returns showing that it becomes close to zero after the first lag. This behaviour is in line with the empirical findings, reinforcing the idea that price changes are not correlated.  

\subsection{Heavy tails}
\label{heavy tails}

\begin{figure}
	\centering  	
	\subfigure[Raw returns distribution. The red line represents a normal distribution superimposed on the return distribution.]{\includegraphics[width=0.49\textwidth]{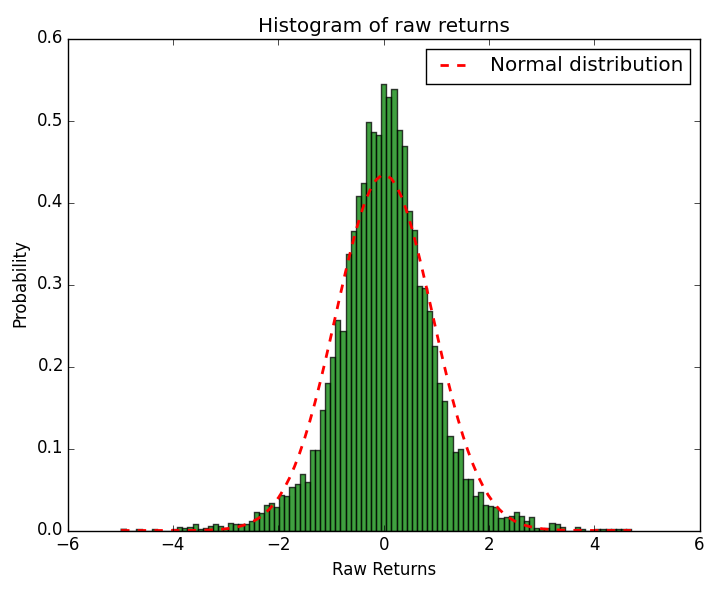}}
	\hfill
	\subfigure[Volatility distribution. The red line represents an exponential distribution superimposed on the volatility distribution.]{\includegraphics[width=0.49\textwidth]{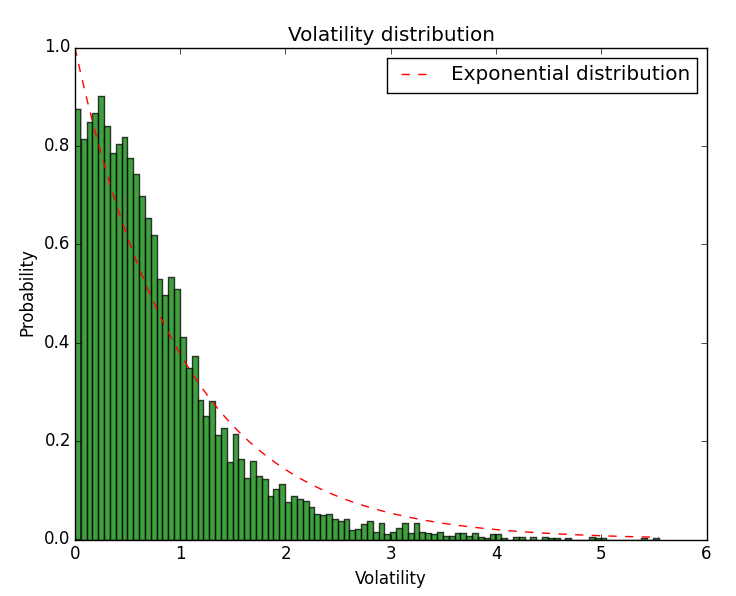}}	
	\caption{Distribution of (a) raw returns and (b) absolute returns.}	
	\label{fig:ret_dist}
\end{figure}

\noindent Another challenging topic in the econometrics literature is the distribution of returns. Consider the distribution on raw returns, plotted in Figure \ref{fig:ret_dist}(a). We can immediately observe that the distribution displays strong deviations from Gaussianity. Returns of stock prices, like returns of many other financial assets, are bell shaped similar to the normal distribution, but contain more mass in the peak and the tail than the Gaussian distribution \citep{fama1965behavior, mandelbrot1963variation}. Such distributions have excess positive kurtosis, being called leptokurtic. Specifically, the simulated return series has an excess kurtosis of 2.49 and a small negative skewness of -0.0055. 

The positive excess kurtosis implies a peakiness of the distribution bigger than normal and a slow asymptotic decay of the probability distribution function. This non-normal decay is coined as heavy (or fat) tail \citep{jansen1991frequency, lux1996stable, gopikrishnan1998inverse, jondeau2003testing, bouchaud2008statistical}. Heavy tails are defined as tails of the distribution that have a higher density than what is predicted under normality assumptions \citep{lebaron2005extreme}. For example, a distribution with exponential decay (as in the normal) is considered thin tailed, while a power decay of the density function is considered a fat tail distribution.  

Next, we consider the distribution of absolute returns, or volatility. As we can see from Figure \ref{fig:ret_dist}(b) the distribution of absolute returns has a bigger decay than the exponential distribution superimposed on it, indicating a heavy tail. The power law behaviour in the distribution of absolute returns can be approximated by fitting a distribution of the form,
\begin{equation}
p(x)\propto x^{-\alpha},
\end{equation}
where $\alpha$ is the power law exponent. Following the work of \cite{alstott2014powerlaw} and the key recommendations from \cite{clauset2009power}, we try to fit a power law distribution to the distribution of absolute returns. Accordingly, we find a power law exponent $\alpha=4.0181$, very similar to the one of empirical data $\alpha_{emp}= 4.031$. These values are computed using the end of the tail that minimizes the differences between the power law and the actual distribution. Note that the power law exponent is in line with most of the other empirical findings reported in \cite{cont2001empirical}. 

Finally, we analyse the distribution of absolute returns by computing the well-known Hill tail index \citep{hill1975simple}. We obtain a Hill index of 3.79 for the simulated absolute returns, which is not significantly different from the empirical Hill index of 3.98 for the S\%P 500 returns. As before, we computed the well-known Hill estimator using the same end of the tail that minimised the differences between the power law and the actual distribution. Notably, most data sets studied reveal a tail index higher that two and less than  five \citep{cont2001empirical}.     

\subsection{Volatility clustering and Long memory}
\label{volatility clustering}

\noindent The absence of autocorrelation discussed in Section \ref{autocorrelation} does not rule out the possibility of nonlinear dependencies of returns. It is well known that absence of serial correlation does not imply independence \citep{pagan1996econometrics}. Even simple visual representations of return series (see Figure \ref{fig:ret_dist}(b)) reveal heteroscedasticity as a violation of the assumption of independently and identically distributed returns. That is, volatility measured as squared or absolute returns is not constant in time. Fluctuations of volatility were first noted by \cite{mandelbrot1963variation} in daily returns of cotton prices. The author reported that periods of high volatility alternate with periods of low volatility. Further early examples of heteroscedasticity were found in \cite{fielitz1971stationarity, wichern1976changes, hsu1977tests}, to name a few.   

Furthermore, nonlinear representations of returns, such as absolute, squared or various powers of returns, exhibit a much higher positive autocorrelation that persists over time. The powers of absolute returns can be seen as measures of volatility, indicating a high degree of predictability of volatility. This phenomenon is stable across different financial instruments and time periods, being a quantitative signature of the well known volatility clustering. That is, large price variations are more likely to be followed by large price variations. This behaviour has been first noted by \cite{mandelbrot1963variation}, while \cite{cont2005long} provides an extensive study on volatility clustering in financial markets. 

We can observe this behaviour just by inspecting the returns generated by the model (Figure \ref{fig:simulation}(c)). Periods of relatively small returns are interrupted by abrupt increases in returns. Moreover, these periods tend to be clustered together. Specifically, small price changes are followed by small ones and large price changes by large ones.

A common way of confirming the presence of volatility clustering is by considering its autocorrelation function. Even though there are different ways of measuring volatility, the most commonly used ones are the absolute returns, 
\begin{equation} \label{eq:20}
C_a(\tau)=corr(|r(t+\tau,\Delta t)|,|r(t, \Delta t)|).
\end{equation}

Plotting out the autocorrelation of volatility measured as absolute returns in Figure \ref{fig:acf}, we observe a positive autocorrelation that persists over time, doubled by its slow decay. This is a clear presence of volatility clustering.

A property closely related to the volatility clustering effect is the decay of the autocorrelation function. The long memory effect specifically addresses this decay. \cite{mandelbrot1971can} was the first one to suggest this stylized fact and observed it in many empirical studies \citep{mandelbrot1979robust}. Long range dependencies have been found across different markets and periods \citep{liu1997correlations, liu1999statistical, cont2005long, chakraborti2011econophysics,cont1997scaling}. Usually, if the decay is slow, similar to a hyperbolic function, we can say that the corresponding process exhibits long memory. A possible explanation of this stylized fact is that investors with different time scales interact in the market, which typically results in a mixture of long-short relaxation times (the delay to secondary reactions unfold), following the impact of exterior events or information. Thus, different relaxation times combine, leading to a hyperbolic decay in the autocorrelation function.  

One way of observing the decay in the autocorrelation function is by fitting a power law of the form, 
\begin{equation}\label{eq:2.32}
C_a(\tau)\sim\frac{A}{\tau^\beta},
\end{equation}
with empirical studies finding the coefficient $\beta\leq 0.5$ \citep{liu1997correlations, cont1997scaling, cont2005long}. We notice that a close fit of the ACF can be obtained, with an exponent $\alpha\approx0.472$ (Figure \ref{fig:acf}).

Another widely used method of testing the long memory effect is by using the Hurst exponent \citep{hurst1951long}, which falls in the range $1/2<H<1$ for a long memory process. Computing the Hurst exponent of the volatility, we obtain the values of 0.70. Similarly, the Hurst exponent in empirical absolute returns is equal to 0.69. Therefore, the presence of a long memory process is clearly demonstrated. 

\subsection{Volume-volatility relations and aggregate Gaussianity}
\label{Volume-volatility}
\begin{figure}
	\centering  	
	\subfigure[Absolute returns vs. Volume]{\includegraphics[width=0.45\textwidth]{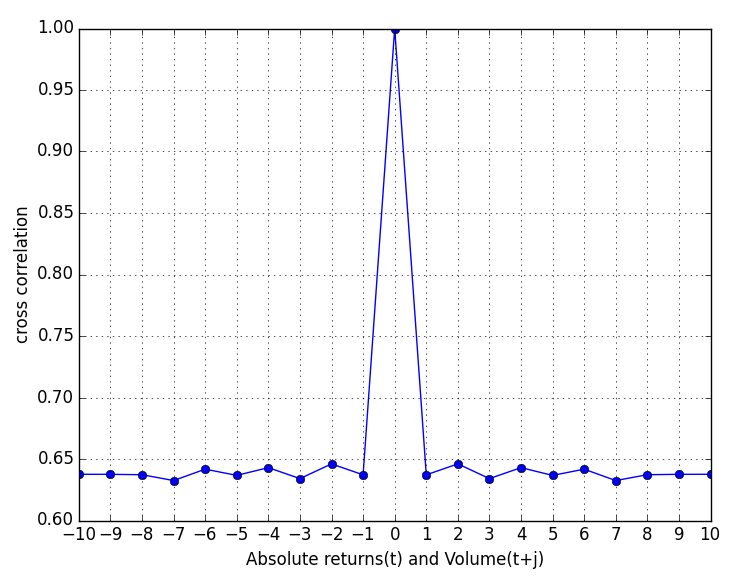}}
	\hfill
	\subfigure[Squared returns vs. Volume]{\includegraphics[width=0.45\textwidth]{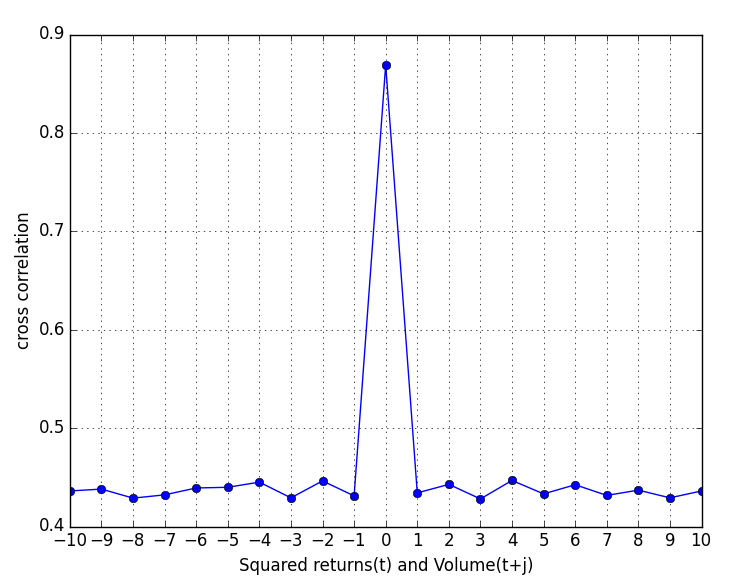}}	
	\caption{Volume-Volatility relations. Cross correlation between volatility measured as (a) absolute returns and (b) squared returns and volume.}	
	\label{fig:cross}
\end{figure}
\noindent The relationship between volatility and volume traded is important for understanding how information is transmitted and embedded in markets. It has been noticed and documented across different financial instruments at different time scales. The volume-volatility relation was first observed by \cite{osborne1959brownian}, with the positive correlation between them being first detected by \cite{ying1966stock}. For an early survey see \cite{karpoff1987relation}. In addition, the causalities of volume on price changes were studied in \cite{chuang2009causality}, concluding that it exhibits a V-shape relation so that the diffusion of volatility distribution increases with volume. 

This stylized fact is at the heart of determining the underlying mechanisms of crashes and rapid changes in the stock prices. Furthermore, an intensive analysis on volume can help investors identify periods in which informational shocks occur, providing valuable information about future price changes. In Figure \ref{fig:cross} we can observe a significantly positive cross correlation between volatility and volume. In this setting, we define the volume at time $t$ as the total absolute demand at $t$ of both fundamentalists and chartists. It is important to mention that all measures of volatility are positively correlated with traded volume. 

The significantly positive cross correlation means that a small (large) trading volume is accompanied by a small (large) change in volatility. This behaviour is reflected in the way future prices are determined by the market maker, demands (and therefore volume) directly influencing volatility changes. It is important to note that the dependence between volume and volatility remains significant as we increase the time lag. From the investor's point of view, this seems to be the only dependency that can be exploited. As we have seen so far, the returns are not correlated (see Section \ref{autocorrelation}) and even though the volatility displays positive correlation, it is less significant and slowly decays as we increase the lag, a signature of volatility clustering (see Section \ref{volatility clustering}).  

A further easily observable property of returns is the aggregate Gaussianity \citep{cont2001empirical}. As we have seen, the distribution of returns has excessive kurtosis and heavy tails, with a higher peak than the normal distribution and power law tails. However, this behaviour changes as we increase the time scale over which the returns are calculated. More specifically, the distribution of returns changes its shape, becoming more like the normal distribution. This stylized fact can easily be observed in different markets and time periods.

\begin{figure}
	\centering  
	\includegraphics[width=0.6\textwidth]{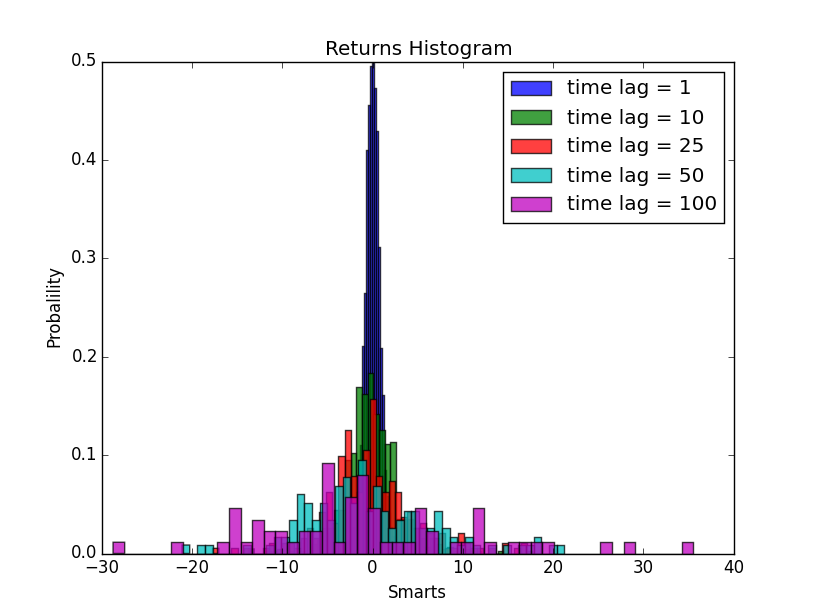}		
	\caption{Aggregate Gaussianity. Distribution of returns calculated at different time scales.}	
	\label{fig:agg}
\end{figure}

In more detail, the aggregate Gaussianity of the simulated returns generated by our model is clearly visible in Figure \ref{fig:agg}. We plot the distribution of returns calculated for different time lags $\tau=1,10,25,50,100$. We can observe how the distribution of returns becomes more like the normal distribution as we increase the time lags. The peaks of the distributions decrease and the tails become wider. Specifically, when returns are calculated at small time scales their distribution is extremely peaked in the middle with very heavy tails. However, as we increase the time scale, the return distributions become more flat, looking like the normal distribution. Finally, as can be seen from Table \ref{table:agg}, the excess kurtosis clearly decreases.

\subsection{Price impact and extreme price events}
\label{Price impact}
\begin{table}
	\centering
	\caption{Values of excess kurtosis and skewness as we increase the time lags at which returns are computed.}
	\label{table:agg}
	\begin{tabular}{|l|l|l|l|l|l|}
		\hline
		time lag & 1       & 10     & 25      & 50      & 100     \\ \hline
		kurtosis & 2.5116  & 2.2022 & 1.6289  &  1.0661  & 0.4954  \\ \hline
		skewness & -0.0049 & -0.0128 & -0.0324 & -0.0409 & -0.0648 \\ \hline
	\end{tabular}
\end{table}

\noindent So far, we have discussed the volume-volatility and return-volatility correlations. The price impact tackles the correlations between the volume and the price changes. More specifically, we are interested in the impact of the volume traded on the prices. There are a few different theories regarding this relation, from linear to non-linear dependencies \citep{kyle1985continuous, toth2011anomalous, moro2009market, gopikrishnan2000statistical,gabaix2003theory}. However, the impact of a single order has been widely found to be a concave function of volume. 

\begin{figure}
	\centering  	
	\subfigure[Price impact]{\includegraphics[width=0.37\textwidth]{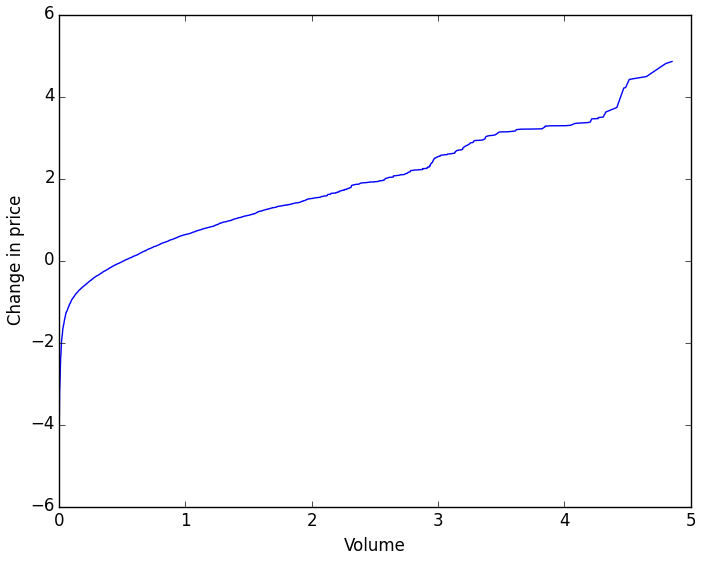}}
	\hfill
	\subfigure[Volume distribution]{\includegraphics[width=0.38\textwidth]{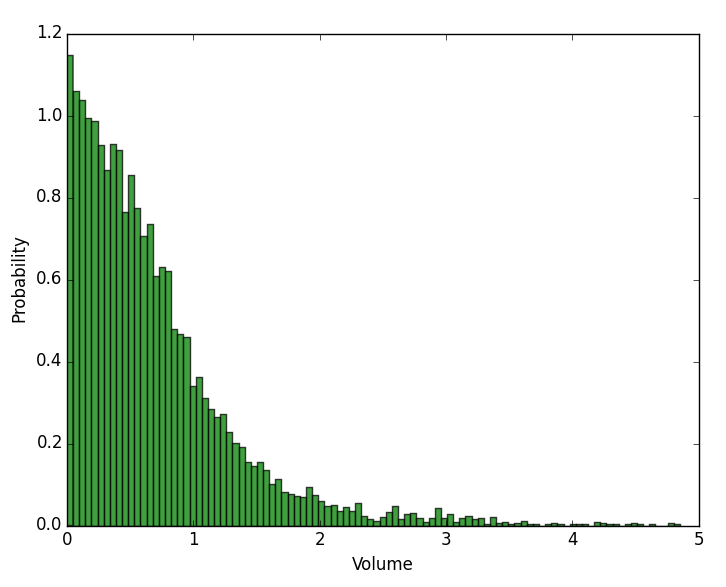}}
	\hfill
	\subfigure[Price series: crash from  $t=5163$ to $t=5175$.]{\includegraphics[width=0.4\textwidth]{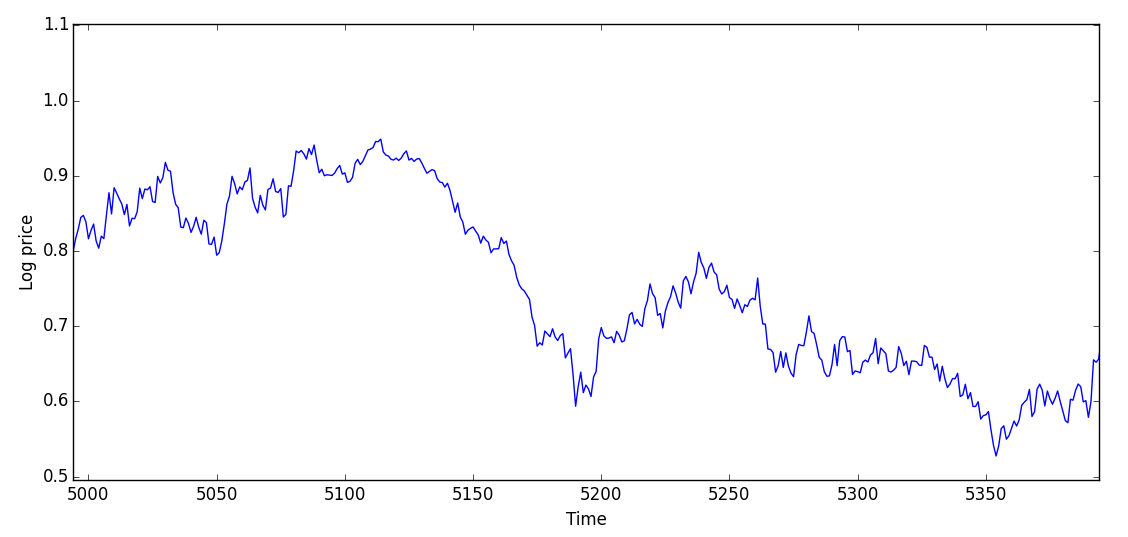}}	
	\hfill
	\subfigure[Majority index: crash from  $t=5163$ to $t=5175$.]{\includegraphics[width=0.45\textwidth]{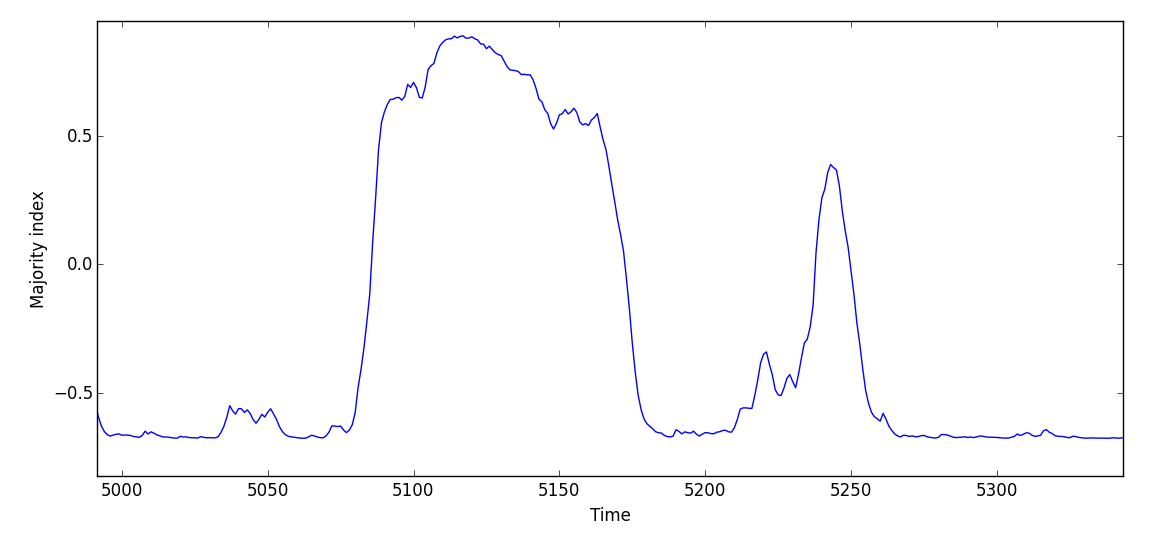}}
	\caption{Price impact function, the volume distribution and an extreme price event.}
	\label{fig:impact}
\end{figure}

Using our simulated data, we plot the changes in prices as a function of volume. As before, we refer to the volume in time period $t$ as the absolute value of the total agents' demand in that specific time period. In Figure \ref{fig:impact}(a), we plot the price changes as a function of volume, clearly obtaining a concave function of price impact. We were expecting this kind of behaviour from the model definition. This is because the market maker computes the next price as a function of the agents' demand multiplied by their market fraction. Clearly, if the absolute demands increase, the next price will also increase, but not linearly.  

Moreover, in Figure \ref{fig:impact}(b), we show the volume distribution. Although the exact distribution of volumes is hard to find, we notice that its tail can be described by a power law with $\alpha\approx3.5$. This is similar to the empirical findings regarding the distribution of the number of trades \citep{gabaix2003theory}.

Finally, we discuss one of the most recently observed phenomena in real financial markets, the so called extreme price events. Following the definition of \cite{johnson2013abrupt}, a spike (crash) is an occurrence of an asset price ticking up (down) at least ten times before ticking down (up) with a price change exceeding 0.8\%. On average, we observe 13 spikes or crashes in each simulation (6866 time periods). In Figure \ref{fig:impact} (c) and (d) we present an example of a crash in prices occurring between periods $t=5163$ and $t=5175$. We observe a drop of $17.19\%$ from the initial price at time $t=5163$ in \ref{fig:impact} (c). There is no surprise that the crash occurs in a period of switch from fundamentalism (majority index $>0.5$) to chartism domination (majority index $<-0.5$), when the herding mechanism suddenly pushes the price away from the fundamental value (\ref{fig:impact} (d)). The price becomes stable when the majority index changes its sign and the fundamentalists begin to increase in numbers and dominate the market. Moreover, we note that even if the changes in prices are significant, the extreme price events do not usually last much longer that 10 time periods. 

\section{Conclusions}
\label{Conclusion}
\noindent Heterogeneous agent-based models that rely on simple trading strategies have proven themselves very efficient in generating important dynamics of real financial markets. Indeed, the fundamentalists vs. chartists models have been shown to successfully capture empirically observed traders' behaviour. In this paper, we described, evaluated and extended one of the most recent and successful in capturing real-life market dynamics. First, we presented its key mechanisms and their particular roles in the model structure. Second, motivated by the violation of a central property of real financial price series, their non-stationarity, we proposed a novel change in the system. In particular, considering the high percentage of stationary price series generated by the model, we change the way the fundamental value is computed and make it follow a geometric Brownian motion. This new model was then shown to overcome the initial violation. 

The main objective of agent-based financial modelling is to propose an alternative to the apparent randomness of financial markets, trying to explain the most important properties of financial data. Specifically, we are interested in simple structures that can reproduce the empirical findings to a high degree and which are quantitatively close to the real ones. To this end, we offer both a quantitatively and a qualitatively analysis of the simulated asset price series, its returns, volume traded and volatility. By providing a wide range of tests and arguments, we demonstrate the presence of a rich set of stylized facts including absence of autocorrelations, heavy tails, volatility clustering, long memory, volume volatility relations, aggregate Gaussianity, price impact and extreme price events. By doing so, our model is the first to match such a rich set of the stylized facts.

The precise numerical values of the model's parameters were obtained following a formal econometric estimation, known as the method of simulated moments. However, we depart from some of the classical examples in the literature where a block bootstrap is used and follow the recent work of \cite{franke2011simple}, thus overcoming the well-known joint-problem associated with older methods. In so doing, this is the first time this more trustworthy alternative estimation has been applied on a model where the interaction between agents is based on a discrete choice approach.

To date, we have discussed a model where the market participants, i.e. fundamentalists and chartists, change their strategies according to a herding mechanism corrected by a price misalignment. However, we are also interested in exploring other empirically backed factors that force the traders to change their beliefs regarding future price movements. In particular, motivated by the growing literature of behavioural finance, more components can be incorporated in the attractiveness level and explore how they alter the interactions between the agents. It would be interesting to investigate how different factors such as gossip, political rumours, or waves of optimism and pessimism can be added in the switching mechanism, making the agents behave in a more realistic way and determine how they change strategies.   
\section*{Acknowledgements}
This work was supported by the EPSRC ORCHID project (EP/1011587/1). The authors acknowledge the use of the IRIDIS High Performance Computing Facility in the completion of this work.

\bibliographystyle{model2-names}
\bibliography{myJournal}







\end{document}